\def\rn{}
\def\nn#1 #2{#2. #1}				
\def\nnn#1 #2 #3{#2. #3. #1}			
\def\nnnn#1 #2 #3 #4{#2. #3. #4 #1}		
\def\nnnnn#1 #2 #3 #4 #5{#2. #3. #4 #5. #1}	
\def\dualand{ and\hbox{ }}				
\def\multiand{, and\hbox{ }}				
\def\rf#1;#2;#3;#4;#5 {{\frenchspacing\par\rn#1, #3 {\bf #4}, #5 (#2). \par}}
\def\rg#1;#2;#3;#4;#5;#6 {{\frenchspacing\par\rn#1, #3 {\bf #4}, #5 (#2). \par}}
\def\rfbook#1;#2;#3;#4;#5 {{\frenchspacing\par\rn#1, {\it #3} (#5, #4, #2).\par}}
\def\rfprep#1;#2;#3 {{\par\frenchspacing\rn#1, #3 (#2).\par}}
\def\rfproc#1;#2;#3;#4;#5;#6 {{\frenchspacing\par\rn#1 #2, in {\it #3}, ed. #4 (#5: #6)\par}}
\def\rfprocp#1;#2;#3;#4;#5;#6;#7 {{\frenchspacing\par\rn#1 #2, in {\it #3}, ed. #4 (#5: #6), p#7\par}}

\def\rg#1;#2;#3;#4;#5;#6 {\par\rn#1 #2, {\it #3}, {\bf #4}, #5 (``#6'') \par}
\def\rf#1;#2;#3;#4;#5 {\par\rn#1, {\it #3}, {\bf #4}, #5 (#2)\par}
\def\rfbook#1;#2;#3;#4;#5 {{\frenchspacing\par\rn#1, {\it #3} (#4: #5, #2)\par}}
\def\rfproc#1;#2;#3;#4;#5;#6 {{\frenchspacing\par\rn#1 #2, in {\it #3}, ed. #4 (#5: #6)\par}}
\def\rfprocp#1;#2;#3;#4;#5;#6;#7 {{\frenchspacing\par\rn#1 #2, in {\it #3}, ed. #4 (#5: #6), p#7\par}}
\def\rfprep#1;#2;#3  {{\par\rn#1, #3, #2\par}}
\def\rfprepp#1;#2;#3 {{\par\rn#1 #2, #3\par}}

\def\Gyr{\>{\rm Gyr}}

\def\etal{{\frenchspacing\it et al.}}
\def\ie{{\frenchspacing\it i.e.}}
\def\eg{{\frenchspacing\it e.g.}}

\def\beq#1{\begin{equation}\label{#1}}
\def\eeq{\end{equation}}
\def\beqa#1{\begin{eqnarray}\label{#1}}
\def\eeqa{\end{eqnarray}}

\def\fig#1{Figure~\ref{#1}}

\def\spose#1{\hbox to 0pt{#1\hss}}
\def\simlt{\mathrel{\spose{\lower 3pt\hbox{$\mathchar"218$}}
     \raise 2.0pt\hbox{$\mathchar"13C$}}}
\def\simgt{\mathrel{\spose{\lower 3pt\hbox{$\mathchar"218$}}
     \raise 2.0pt\hbox{$\mathchar"13E$}}}
\def\simpropto{\mathrel{\spose{\lower 3pt\hbox{$\mathchar"218$}}
     \raise 2.0pt\hbox{$\propto$}}}

\def\ed{\end{document}}

\documentclass[twocolumn,amsmath,nofootinbib]{revtex4} 
\usepackage{url}
\begin{document}
\input{epsf.sty}




\def\affilmrk#1{$^{#1}$}
\def\affilmk#1#2{$^{#1}$#2;}

\title{How Unlikely is a Doomsday Catastrophe?}

\author{Max Tegmark$^1$ \& Nick Bostrom$^2$ }
\address{
$^1$Department of Physics, Massachusetts Institute of Technology, Cambridge, MA 02139, USA\\
}
\address{
$^2$Future of Humanity Institute, Faculty of Philosophy, Oxford University, OX14JJ, Oxford, UK\\
}

\date{
December 18, 2005.
This paper is an extended version of the {\it Brief Communication} published
in {\it Nature}, {\bf 438}, 754 \cite{risk}.}

\begin{abstract}
Numerous Earth-destroying doomsday scenarios have recently been analyzed, 
including breakdown of a metastable vacuum state and planetary destruction 
triggered by a ``strangelet'' or microscopic black hole.
We point out that many previous bounds on their frequency 
give a false sense of security: one
cannot infer that such events are rare from the the fact that Earth has survived 
for so long, because observers are by definition in places lucky enough to have avoided destruction.
We derive a new upper bound of one per $10^9$ years (99.9\% c.l.) on the exogenous terminal catastrophe rate 
that is free of such selection bias, using planetary age distributions and the relatively late formation
time of Earth.
\end{abstract}


  
\maketitle


\setcounter{footnote}{0}

\section{Introduction}
\label{IntroSec}

As if we humans did not have enough to worry about, scientists have recently highlighted catastrophic scenarios that
could destroy not only our civilization, but perhaps even our planet or our entire observable universe.
For instance, fears that heavy ion collisions at the Brookhaven
Relativistic Heavy Ion Collider (RHIC) might initiate such a catastrophic process
triggered a detailed technical report on the subject \cite{Jaffe00}, focusing on three risk categories:
\begin{enumerate}
\item Initiation of a transition to a lower vacuum state, which would propagate
outward from its source at the speed of light, possibly destroying the universe as we know it \cite{Frampton76,Hut83,Jaffe00}. 
\item Formation of a black hole or gravitational singularity that accretes ordinary matter, 
possibly destroying Earth \cite{Hut83,Jaffe00}.
\item Formation of a stable ``strangelet'' that accretes ordinary matter and converts it to strange matter, 
possibly destroying Earth \cite{Dar99,Jaffe00}.
\end{enumerate}
Other catastrophe scenarios range from uncontroversial to highly speculative:
\begin{enumerate}
\setcounter{enumi}{3}
\item Massive asteroid impacts, nearby supernova explosions and/or gamma-ray bursts, potentially sterilizing Earth.
\item Annihilation by a hostile space-colonizing robot race.
\end{enumerate}
The Brookhaven report \cite{Jaffe00} concluded that if 1-3 are possible, then they will with overwhelming likelihood be
triggered not by RHIC, but by naturally occurring high-energy astrophysical events such as cosmic ray collisions.
Risks 1-5 should probably all be treated as {\it exogenous}, \ie, uncorrelated with human activities and 
our technical level of development.
The purpose of the present paper is to assess the likelihood per unit time of exogenous 
catastrophic scenarios such as 1-5.

One might think that since life here on Earth has survived for nearly 4 Gyr (Gigayears), 
such catastrophic events must be extremely rare.
Unfortunately, such an argument is flawed, giving us a false sense of security.
It fails to take into account the observation selection effect \cite{Carter74,BostromBook} that precludes any
observer from observing anything other than that their own species has survived up to the point where they make
the observation. Even if the frequency of cosmic catastrophes were very high, we should still expect to find
ourselves on a planet that had not yet been destroyed. The fact that we are still alive does not even seem to
rule out the hypothesis that the average cosmic neighborhood is typically sterilized by vacuum decay, say, every 10000 years, 
and that our own planet has just been extremely lucky up until now. If this hypothesis were true, future
prospects would be bleak.

We propose a way to derive an upper bound on cosmic catastrophe frequency that is
unbiased by such observer selection effects. 
We argue that planetary and stellar age distributions bound the rates of many doomsday scenarios, and that scenarios evading this bound 
(notably vacuum decay) are instead constrained by the relatively late formation time of Earth.
The idea is
that if catastrophes were very frequent, then almost all intelligent civilizations would arise much earlier than
ours did. Using data on planet formation rates, it is possible to calculate the distribution of birth dates for
intelligent species under different assumptions about the rate of cosmic sterilization. Combining this with the
information about our own temporal location enables us to conclude 
that the cosmic sterilization rate
for a habitable planet is at most of order one per Gigayear.

\begin{figure*} 
\vskip-7.0cm
\centerline{\epsfxsize=15cm\epsffile{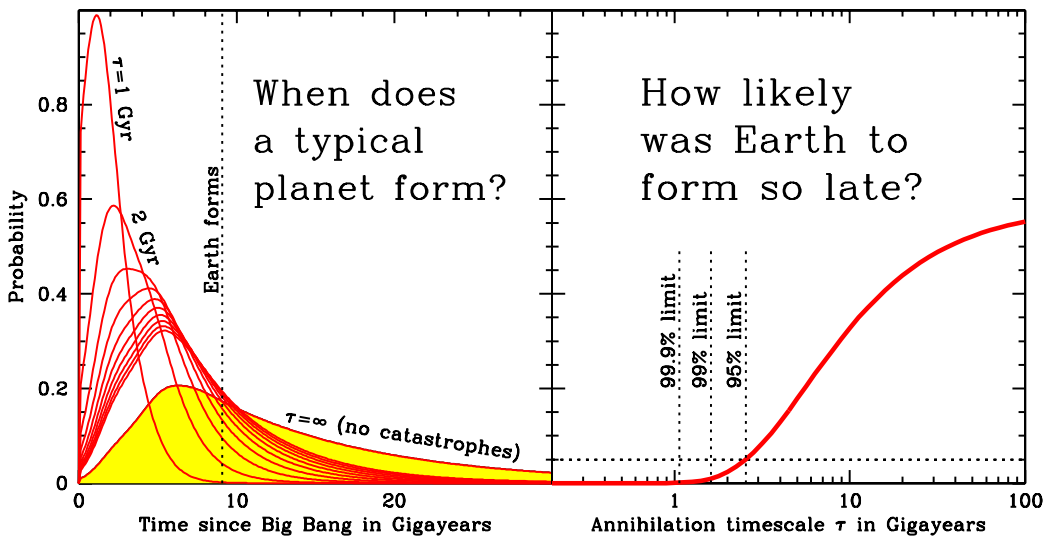}}
\vskip-1cm
\caption[1]{\label{doubleFig}\footnotesize%
The left panel shows the probability distribution for observed planet formation time
assuming catastrophe timescales $\tau$ of $\infty$ (shaded), 10, 9, 8, 7, 6, 5, 4, 3, 2 and 1 Gyr, respectively
(from right to left).
The right panel shows the probability of observing a formation time $\ge$9.1 Gyr (that for Earth), 
\ie, the area to the right of the dotted line in the left panel. 
}
\end{figure*}

\section{An upper bound on the catastrophe rate}

Suppose planets get randomly sterilized or destroyed at some rate $\tau^{-1}$ which we will now constrain.
This means that the probability of a planet surviving a time $t$ decays exponentially, as $e^{-t/\tau}$. 

The most straightforward way of eliminating observer selection bias is to use only information 
from objects whose destruction would not yet have affected life on Earth.
We know that no planets from Mercury to Neptune in our solar system have been 
converted to black holes or blobs of strange matter during the past 4.6 Gyr, since their masses would still be detectable via their gravitational
perturbations of the orbits of other planets.
This implies that the destruction timescale $\tau$ must be correspondingly large --- unless their destruction is be linked to ours, 
either by a common cause or by their implosion resulting in the emission of doomsday particles like black holes or strangelets that would in turn destroy Earth.
This observer selection effect 
loophole is tightened if we consider extrasolar planets that have been seen to partially eclipse their parent star \cite{Pont05} 
and are therefore known 
not to have imploded. The doomsday particles discussed in the literature would be more readily captured gravitationally by a star than by a planet, 
in which case the observed abundance of very old ($\simgt 10$ Gyr) stars (\eg, \cite{Hansen02}) would further sharpen the lower bound on $\tau$.

The one disaster scenario that exploits the remaining observer bias loophole and evades all these constraints is vacuum decay,
either spontaneous or triggered by a high-energy event. 
Since the bubble of destruction expands with the speed of light, we are prevented from observing the destruction of other objects: 
we only see their destruction at the instant when we ourselves get destroyed.
In contrast, if scenarios 2 or 3 involved doomsday particle emission and proceed as a chain reaction spreading subluminally,
we would observe spherical dark regions created by expanding destruction fronts that have not yet reached us.
%
We will now show that the vacuum decay timescale can be bounded by a different argument.

The formation rate $f_p(t_p)$ of habitable planets as a function of time since the Big Bang
is shown in \fig{doubleFig} (left panel, shaded distribution). This estimate is from
\cite{Lineweaver04}, based on simulations including the effects of heavy element buildup,
supernova explosions and gamma-ray bursts.
If regions of space 
get randomly sterilized or destroyed at a rate $\tau^{-1}$, then the probability 
that a random spatial region remains unscathed decays as $e^{-t/\tau}$. 
This implies that the conditional probability distribution $f_p^*(t_p)$
for the planet formation time $t_p$ seen by an observer
is simply the shaded distribution $f_p(t_p)$ multiplied by $e^{-t_p/\tau}$ and rescaled to integrate
to unity, giving the additional curves in \fig{doubleFig} (left panel).\footnote{ 
Proof: Let $f_o(t_o)$ denote the probability distribution for the time $t_o$ after planet formation 
when an observer measures $t_p$. In our case, $t_o=4.6$ Gyr.
We obviously know very little about this function $f_o$, but it fortunately drops out of our calculation.
The conditional probability distribution for $t_p$, marginalized over $t_o$, is
\beq{fpEq}
f_p^*(t_p)\propto\int_0^\infty f_o(t_o)f_p(t_p) e^{-{t_o+t_p\over\tau}}dt_o
\propto f_p(t_p) e^{-{t_p\over\tau}},
\eeq
independently of the unknown distribution $f_o(t_o)$,
since $e^{-(t_o+t_p)/\tau}=e^{-t_o/\tau}e^{-t_p/\tau}$ and hence the entire integrand
is separable into a factor depending on $t_p$ and a factor depending on $t_o$.
}
As we lower the catastrophe timescale $\tau$, the resulting distributions (left panel)
are seen to peak further to the left and the probability that Earth formed as late
as observed (9.1 Gyr after the Big Bang) or later drops (right panel).
The dotted lines show that we can rule out 
the hypothesis that $\tau<2.5\Gyr$ at 95\% confidence, and that 
the corresponding 99\% and 99.9\% confidence limits are 
$\tau>1.6\Gyr$ and $\tau>1.1\Gyr$, respectively.

Risk category 4 is unique in that we have good direct measurements of the frequency 
of impacts, supernovae and gamma-ray bursts that are free from observer selection effects
Our analysis therefore used the habitable planet statistics from \cite{Lineweaver04}
that folded in such category 4 measurements.


Our bounds do not apply in general to disasters of anthropogenic origin, such as ones
that become possible only after certain technologies have been developed,
\eg, nuclear annihilation or extinction via engineered microorganisms or nanotechnology.
Nor do they apply to natural catastrophes that would not permanently destroy or sterilize a planet. 
In other words, we still have 
plenty 
to worry about \cite{Leslie96,Bostrom02,Rees03,Posner04}.
However, our bounds do
apply to exogenous catastrophes (\eg, spontaneous or cosmic ray triggered ones) whose frequency is uncorrelated with 
human activities, as long as they cause permanent sterilization. 

 
Our numerical calculations made a number of assumptions.
For instance, we treated the exogenous catastrophe rate $\tau^{-1}$
as constant, even though one could easily imagine it varying by of order 
10\% over the relevant timescale, since our bound on $\tau$ is about 10\% of the
age of the Universe.\footnote{As pointed out by Jordi Miralda-Escude (private communication), 
the constraint from vacuum decay triggered by bubble nucleation is even stronger than our conservative estimate.
The probability that a given point is not in a decayed domain at time $t$ is the probability of no bubble nucleations
in its backward light cone, whose spacetime 4-volume $\propto t^4$ for both matter-dominated and radiation-dominated expansion.
A constant nucleation rate per unit volume per unit time therefore gives 
a survival probability $e^{-(t/\tau)^4}$ for some destruction timescale $\tau$.
Repeating our analysis with $e^{-t/\tau}$ replaced by the sharper cutoff $e^{-(t/\tau)^4}$ sharpens our constraint.
Our quoted bound corresponds to the conservative case where $\tau$ greatly exceeds the age of the universe at the dark energy domination
epoch, which gives a backward lightcone volume $\propto t$.
} 
Second, the habitable planet formation rate involved several assumptions detailed in \cite{Lineweaver04}
which could readily modulate the results by 20\%. 
Third, the risk from events triggered by cosmic rays will vary slightly with location if the cosmic ray rate does.
Fourth, due to cosmological mass density fluctuations, the mass to scatter off of varies by about 10\% from one 
region of size $c\tau\sim 10^9$ lightyear region to another,
so the risk of cosmic-ray triggered vacuum decay will vary on the same order.

In summary, although a more detailed calculation could change the quantitative bounds by a factor of order unity, 
our basic result that the exogenous extinction rate is tiny on human and even geological timescales appears rather robust.

\section{Conclusions}

We have shown that life on our planet is highly unlikely to be annihilated by an 
exogenous catastrophe during the next $10^9$ years.
This numerical limit comes from the scenario on 
on which we have the weakest constraints: vacuum decay, constrained only by 
the relatively late formation time of Earth.
%
conclusion also translates into a bound on hypothetical anthropogenic disasters
caused by high-energy particle accelerators (risks 1-3).

This holds because the occurrence of exogenous catastrophes,
\eg, resulting from cosmic ray collisions, places an upper bound on the frequency of
their anthropogenic counterparts. Hence our result closes the logical loophole of selection bias and 
gives reassurance that the risk of accelerator-triggered doomsday is extremely small, 
so long as events equivalent to those in our experiments occur more frequently in 
the natural environment.
Specifically, the Brookhaven Report \cite{Jaffe00} suggests that possible disasters would be triggered 
at a rate that is at the very least $10^3$ times higher for naturally occurring events than for high-energy particle accelerators.
Assuming that this is correct, our 1 Gyr limit therefore translates into a conservative upper bound of $1/10^3\times 10^9=10^{-12}$
on the annual risk from accelerators, which is reassuringly small.


%
%

\clearpage
{\bf Acknowledgements:} 

The authors are grateful to 
Adrian Kent, Jordi Miralda-Escude and Frank Zimmermann for spotting loopholes in the first version of this paper,
to the authors of \cite{Lineweaver04} for
use of their data, and to Milan Circovic, Hunter Monroe, John Leslie, Rainer Plaga and Martin Rees for helpful comments and discussions,
Thanks to 
Paul Davies, Charles Harper, Andrei Linde and the 
John Templeton Foundation for organizing a workshop where this work was initiated.
This work was supported by NASA grant NAG5-11099,
NSF CAREER grant AST-0134999, and fellowships from the David and Lucile
Packard Foundation and the Research Corporation.  


\end{document}